# Paramagnetic phase of $MnTe_2$: complete one phonon Raman spectrum

Taras Palasyuk[1,*], Dominik Kurzydłowski[2], Ashutosh S. Wadge[3], Rajmund Bacewicz[1], Madhavi H. Dalsaniya[4], Michał Wierzbicki[1], Daniel Jastrzebski[3], Andrzej Wiśniewski[3], Wojciech Szuszkiewicz[5], Igor Antoniazzi[6], Bruno Camargo[6], Maciej R. Molas[6], Adam Babiński[6], Krzysztof Zberecki[1] and Cezariusz Jastrzebski[1]

*[1] Faculty of Physics, Warsaw University of Technology, Koszykowa 75, 00-662 Warsaw, Poland*
*[2] Faculty of Mathematics and Natural Sciences, Cardinal Stefan Wyszynski University in Warsaw, Wóycickiego 1/3, 01-938 Warsaw, Poland*
*[3] International Research Centre MagTop, Institute of Physics, Polish Academy of Sciences, Aleja Lotnikow 32/46, 02-668 Warsaw, Poland*
*[4] Faculty of Mechanical Engineering, Bialystok University of Technology, Wiejska 45C, 15-351 Bialystok, Poland*
*[5] Institute of Physics, Polish Academy of Sciences, Aleja Lotnikow 32/46, 02-668 Warsaw, Poland*
*[6] Faculty of Physics, University of Warsaw, Pasteura 5, 02-093 Warsaw, Poland*

*Corresponding author:
Taras Palasyuk taras.palasyuk@pw.edu.pl

**ABSTRACT**

In this study we present a complete first-order Raman scattering spectrum of paramagnetic phase of $MnTe_2$ – a material, where new type of spin texture induced by noncoplanar antiferromagnetic ordering below the Néel temperature ($T_N \approx 90$ K) gives rise to unconventional scenario of magnetism. Basic properties of all five phonon modes ($A_g$, $E_g$ and three $T_g$) predicted by the group theory have been successfully approached through polarization, excitation and temperature dependent measurements. In particular, the ever-missing phonon mode of $T_g$ symmetry has been unambiguously resolved in the room temperature Raman spectrum at approximately 107 $cm^{-1}$, just 1.5 $cm^{-1}$ higher in frequency compared to the $E_g$ mode. Density functional theory calculations reproduce the observed Raman-active phonon spectrum and support the assignment of all five first-order Raman modes, including the near-degeneracy of the $E_g$ and one of the $T_g$ vibrations. The calculations further reveal a dense manifold of low-energy antiferromagnetic spin configurations, separated from the ground state by only a few meV per formula unit. Furthermore, strong signals of inelastically scattered light from the paramagnetic (PM) phase of $MnTe_2$ were acquired with long wavelength excitations, i.e. 633 nm and 785 nm, revealing the $2A_g$ overtone at approx. 360 $cm^{-1}$. The enhanced absorption of the light in the near infrared range could be a putative reason behind the anomalous angular dependence of scattered intensity of the highest frequency $T_g$ phonon mode (183 $cm^{-1}$) measured in PM $MnTe_2$ with parallelly aligned polarizers. Finally, upon cooling just below the $T_N$, the measured properties of phonons remained basically unperturbed by the incipient long-range antiferromagnetic ordering.

## I. INTRODUCTION

Unconventional magnetism such as altermagnetism, noncollinear spin-split antiferromagnetism, and anomalous Hall antiferromagnetism has attracted considerable attention due to its fundamental scientific relevance and promising potential for spintronic applications (e.g. [1], [2], [3], [4]). These magnetic phases extend the conventional framework of ferromagnetism and antiferromagnetsim by introducing symmetry-driven spin textures that can generate unconventional transport responses without relying on relativistic interactions such as spin–orbit coupling (SOC). Altermagnets, in this respect, are particularly compelling class of materials, characterized by collinear antiferromagnetic ordering and spin-split electronic bands. In these systems, the interplay between crystal and time-reversal symmetries allows for spin-polarized transport while preserving a vanishing net magnetization. Theoretical symmetry analyses have already predicted a broad landscape of candidate materials [5], although only a few altermagnets have been experimentally verified (e.g., MnTe [6], CrSb [7], $RuO_2$ [8]), underscoring the challenges in realizing and identifying these phases.

In parallel with these developments, $MnTe_2$ has emerged as a fascinating material hosting a different flavor of unconventional magnetism, showing the emergence of a new type of nonrelativistic spin splitting in a noncolinear antiferromagnetic order, which has recently received direct experimental confirmation by Zhu et al. [9] in their spin- and angle-resolved photoemission spectroscopy (SARPES) measurements. In this compound the in-plane components of spin are found to be antisymmetric about the high-symmetry planes of the Brillouin zone, comprising a plaid-like spin texture in the antiferromagnetic ground state. Such spin pattern originates from the intrinsic AFM order instead of spin–orbit coupling and diminishes at the high-temperature paramagnetic state [9].

$MnTe_2$ crystallizes in the pyrite structure described by the non-symmorphic space group $T_h^6$ ($Pa\overline{3}$ (nr 205)) [10]. The cubic unit cell contains 12 atoms (four formula units), wherein the manganese cations ($Mn^{2+}$) occupy the sites of a face centered cubic sublattice and the tellurium species are arranged in dumb-bells pairs (diatomic anions) aligned along the four equivalent [111] directions.

Raman spectroscopy, an optical technique, dealing with light inelastically scattered in materials, provides essential information on changes in the dynamics of crystalline lattice induced by various collective excitations (quasiparticles) in material such as phonons and magnons. Group theoretical analysis predicts five zone-center phonon modes that are active in the first-order Raman spectrum (e.g., [11], [12], [13]). Owing to the centrosymmetric positions occupied by the manganese atoms, all Raman-active vibrations involve exclusively displacements of tellurium atoms. These comprise one totally symmetric mode ($A_g$), one doubly degenerate mode ($E_g$) and three triply degenerate modes ($T_g$). The $A_g$ and $E_g$ correspond to a purely stretching and librational motions of Te-Te pairs (displacing along and perpendicular to the bond axes, respectively) whereas $T_g$ modes are combinations of stretching (in- and out-of-phase) and librational motions (please refer to Supplementary **Figure S5**).

To the best of our knowledge, ambient-pressure Raman spectra of pyrite-type $MnTe_2$ have been reported in only two peer-reviewed studies [14], [15]. As noted by the authors of both works, the Raman features of $MnTe_2$ were substantially obscured by intense signals originating from crystalline elemental tellurium (hereafter denoted as c-Te). The presence of significant amounts of elemental tellurium was attributed to partial laser-induced decomposition of the $MnTe_2$ samples, which was reported to occur even at temperatures as low as 90 K [14]. Consequently, only two of the five Raman-active phonon modes could be assigned unambiguously, namely the bands observed at 175 and 180 $cm^{-1}$, corresponding to $A_g$ and $T_g^{(2)}$ Te-Te symmetric and asymmetric stretching modes, respectively [14]. In the same study, the frequency of the purely librational mode, $E_g$, was estimated to lie in the range of 126 – 128 $cm^{-1}$ [14]. However, our results indicate that this assignment is not correct, as discussed below.

At ambient pressure, for a number of metal di-chalcogenides compounds ($MX_2$) of pyrite structure all five Raman-active modes were clearly identified in the measured spectra: $OsS_2$, $RuS_2$, $RuSe_2$ [16], $RuTe_2$, $OsTe_2$ [14], $MnS_2$ [13], $MnSe_2$ [14], $FeS_2$, $CoS_2$ [17]. At the same time only four Raman-allowed were observed for the following $MX_2$: $CuS_2$ [18], studied at ambient pressure and for the pyrite phases stable at high-pressure conditions. In the case of $NiSe_2$ [19], there is one report revealing four modes in the ambient Raman spectrum while in the study of ElGhazali et al. [20] all five Raman modes were clearly observed at pressure 0.7 GPa for $NiS_2$ [21] bringing about certain controversy. While preparing this manuscript we became aware of Y. Li's PhD Thesis (2024) [22] and two unpublished manuscripts available through the arXiv preprint repository [23], [24], which among others include analyses of the Raman spectrum of $MnTe_2$. They reveal the latest update on this topic, albeit the presented data should be viewed as preliminary at this stage. The first two works investigated $MnTe_2$ as a major phase in the analyzed samples [22], [23], however, properties of the paramagnetic phase were probed only at 120 K. Owing to the fact that the samples did not contain considerable amount of elementary tellurium, it was possible to identify the lowest frequency $T_g$ phonon mode at about 98 $cm^{-1}$ (1.6 K) [22]. At the same time, overall weak scattering intensity precluded from unambiguous assigning the rest of the modes detected at 107 $cm^{-1}$, 127 $cm^{-1}$ and 183 $cm^{-1}$ (1.6 K) [22]. In the most recent preprint, $MnTe_2$ constituted only 1.6 % of the analyzed sample and was present as a secondary phase alongside the major constituent MnTe (manganese monotelluride) [24]. Nevertheless, the authors reported DFT-calculated frequencies for all five Raman-active phonon modes based on the ground-state AFM structure. As in the earlier studies [14], however, the presence of elemental tellurium (0.5%) prevented a reliable analysis of the Raman features below 150 $cm^{-1}$. Yet another preprint reveals that possible $MnTe_2$ impurities, even at low concentrations, severely complicate the interpretation of MnTe's lattice dynamics, which to date are still not fully understood [25].

Hence, $MnTe_2$ stands out among the family of transition metals dichalcogenides ($MX_2$) for the knowledge on its basic phonon properties, especially in its paramagnetic phase, remains elusive to the date. Here we present a comprehensive Raman spectroscopy investigation of paramagnetic phase of $MnTe_2$ covering the whole temperature range of its stability, i.e. 80 – 340 K (Néel temperature ($T_N$) = 85 ÷ 90 K [22], [26] and ref. therein, this report). This study settles all so far disputable issues related to the basic properties of the whole set of Raman-active phonon modes of paramagnetic phase of $MnTe_2$. In particular, all three modes with $T_g$ symmetry have been unambiguously identified. Our investigation is complemented by ab initio assessment $MnTe_2$ lattice modes in its antiferromagnetic phase revealing similar sequence of phonon modes. Furthermore, it has been observed that the non-zero angle dependence measured for the highest $T_g$ mode as well as significantly increased intensity of $E_g$ and $A_g$ are consequences of enhanced absorption of $MnTe_2$ in the long-wavelength range of spectrum. Also, our study elucidates the appearance of signals related to elementary Te. The origin of signals related to c-Te has been accounted for by the laser induced crystallization of the amorphous tellurium (a-Te).

The remainder of this manuscript is organized as follows. Section II presents and discusses experimental results, including polarization-resolved, excitation-dependent, and temperature-dependent Raman measurements. The results of the theoretical calculations are presented in Section III and are subsequently compared with the experimental observations. Section IV summarizes the main findings and conclusions of the study, while Section V describes the experimental procedures and computational methodology employed in this work.

## II. RESULTS AND DISCUSSION

In this study measurements were aimed at gaining insights into angular, temperature and excitation dependences of the Raman spectra in $MnTe_2$. Polarization and temperature dependent measurements were performed using laser line 633 nm (1.96 eV) while for the energy dependent measurements additional laser lines 488 nm (2.54 eV), 514.5 nm (2.41 eV) and 785 nm (1.58 eV) were utilized. All measurements were performed in backscattering geometry.

### Polarization-resolved measurements

The incident light was polarized by a linear polarizer ("Polarizer") always horizontally along directions within the YZ plane of the probed $MnTe_2$ crystal, as shown in **Figure 1(a)**. Inelastically backscattered light (Raman signal) was analyzed by another linear polarizer ("Analyzer") placed at the entrance of the spectrometer. Measurements were performed for two positions of the analyzer: parallel and perpendicular to polarization vector of the incident light (**Figure 1(b)**), which is equivalent to other conventions known as collinear/crossed and VV/VH configurations, respectively. Raman spectra were collected from (100) crystal face of the $MnTe_2$ sample. Polarization angle was defined as an angle between the incident polarization vector and crystal axis Z. Its value was incrementally changed by rotating the sample with respect to the incident laser within YZ plane. Hence pure parallel and perpendicular configurations were realized when value of the polarization angle was set to $0^o$ and $90^o$ degree, respectively. These two ultimate cases correspond to $X(ZZ)\bar{X}$ and $X(YZ)\bar{X}$ measurement configurations according to Porto notation.

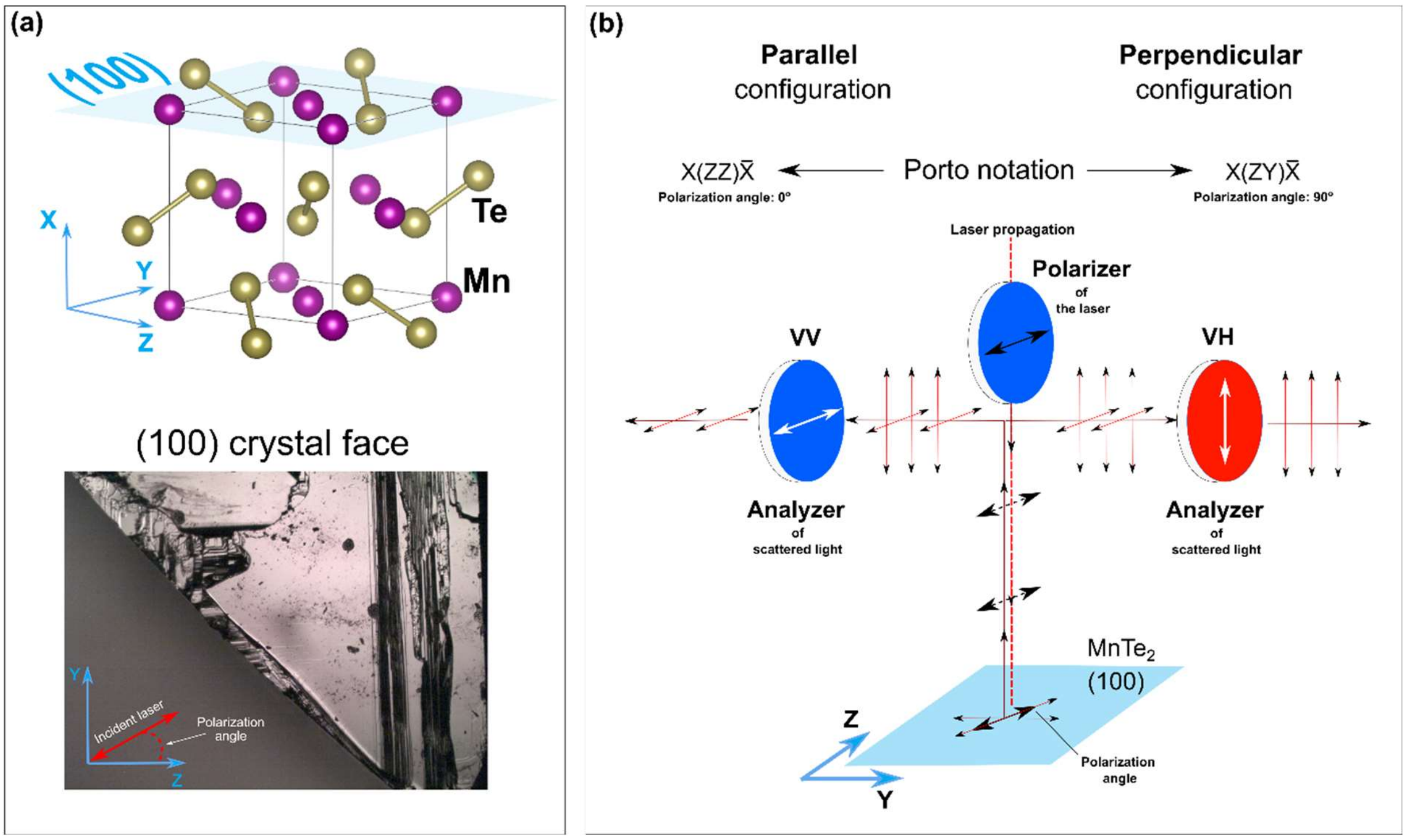

**Figure 1. (a)** schematic view of the unit cell of the pyrite structure $MnTe_2$ with the (100) plane highlighted in blue and optical microscope image of the probed crystal face (100) of sample showing the crystallographic axes and indicating the angle between the polarization direction of the incident light and the crystal Z axis. **(b)** Parallel (VV) and perpendicular (VH) experimental layouts used for measurements

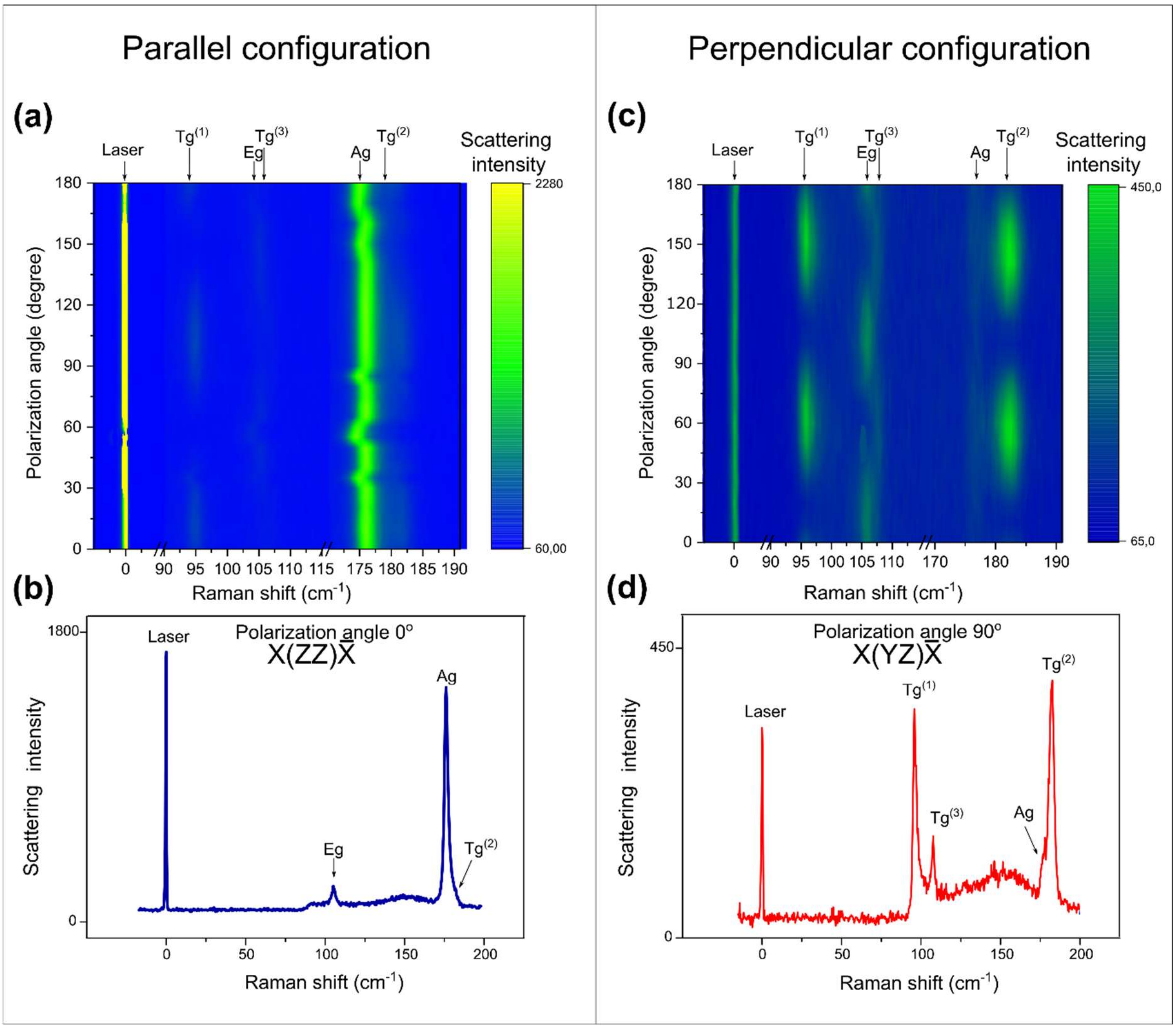

**Figure 2.** Representative results of room temperature polarized Raman measurements of $MnTe_2$ recorded with 633 nm (1.96 eV) laser line. **(a)**, **(b)** contour plots corresponding to angular dependences of the Raman intensity spectra recorded at parallel and cross-polarization configurations, respectively. **(c)**, **(d)** Representative Raman spectra acquired at parallel and cross-polarization configurations, respectively.

In **Figure 2**, representative results of the polarization-resolved measurements are presented. The symmetries of all five Raman-active phonon modes were unambiguously determined from the angular dependence of their scattering intensities, as illustrated by contour and polar plots shown in **Figure 2** and **Figure 3**, respectively. . In order of increasing frequency, the observed modes were assigned to $T_g^{(1)}$, $E_g$, $T_g^{(3)}$, $A_g$ and $T_g^{(2)}$. The nomenclature adopted for the $T_g$ modes follows the convention introduced by Kleppe and [12].

The two-dimensional contour plots of the scattered intensity collected over the polarization-angle range of 0°–180°, together with the Raman spectra recorded in parallel and perpendicular polarization configurations, unambiguously reveal the presence of an additional mode of Tg symmetry, denoted as $T_g^{(3)}$, in close proximity to the $E_g$ mode The frequency separation between the $E_g$ and $T_g^{(3)}$ modes is only approximately 1.5 cm$^{-1}$. Consequently, the $T_g^{(3)}$ mode may remain unresolved in conventional Raman measurements lacking polarization discrimination or in spectra recorded exclusively in the parallel polarization configuration.

In addition to the well-defined phonon modes, the Raman spectra exhibit a broad feature centered at approximately 150 cm$^{-1}$. Based on previous reports, this feature is attributed to amorphous tellurium [27], [28].

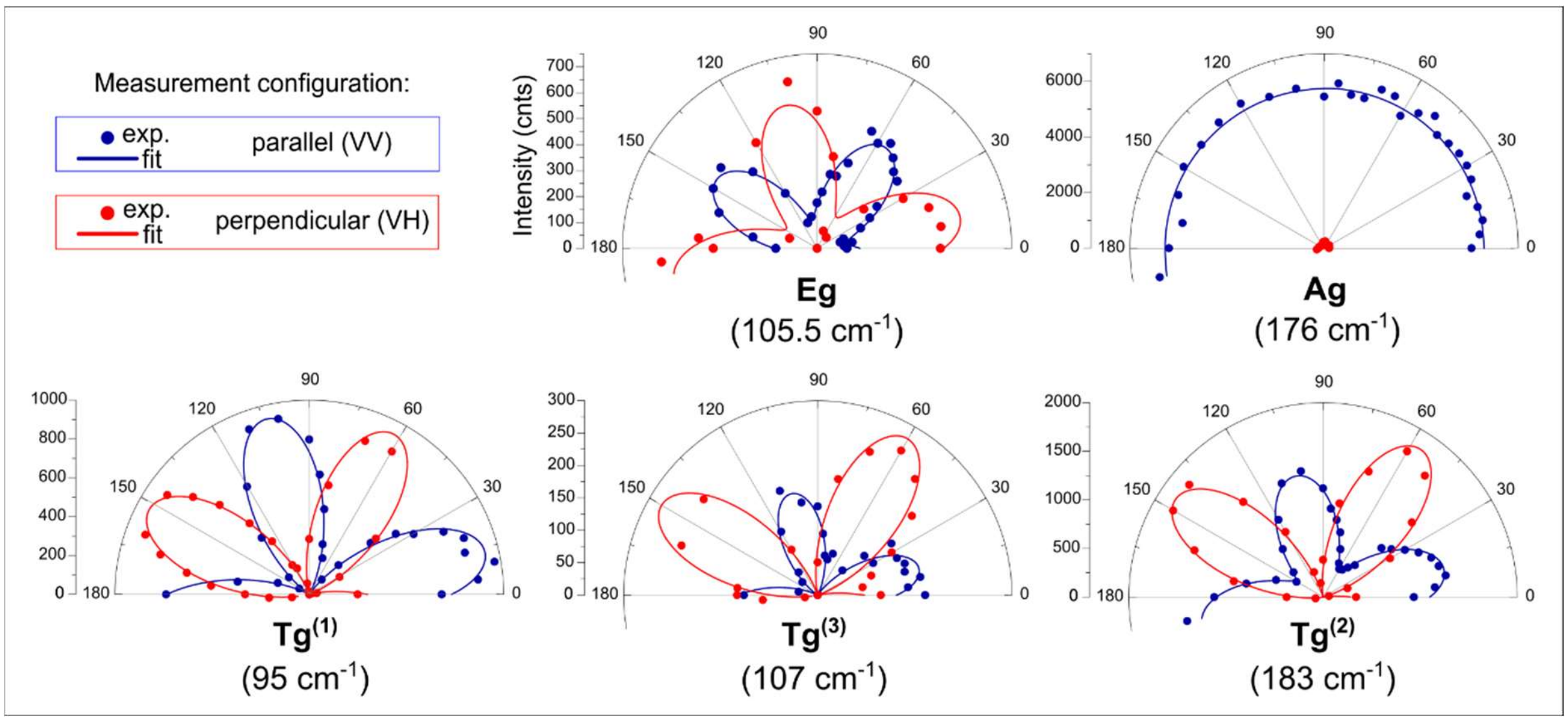

**Figure 3.** Angular dependence plots for $E_g$, $A_g$, $T_g^{(1)}$, $T_g^{(3)}$ and $T_g^{(2)}$ phonon modes. Data acquired for parallel and perpendicular alignments of polarizer and analyzer are shown as blue and red plots, respectively. Experimental data sets (solid circles) were fit by solid lines derived from analytical expressions listed in **Table 1**.

**Table 1.** Raman intensity as a function of the angle between incident and scattered light polarization. Tensor elements '*d*' and '*e*' are assumed to be equal (*d* = *e*) in Raman tensor of $T_g$ modes. C – additive constant (C = 0 for $T_g^{(1)}$ and $T_g^{(3)}$ modes, C $\neq$ 0 for the $T_g^{(2)}$ mode); $\varphi$ - offset angle.

| **Mode** | Raman tensors [13] (**R**) | **Intensity ($I \propto |\hat{e}_i \cdot R \cdot \hat{e}_s|^2$),** **$\hat{e}_i$, $\hat{e}_s$** – polarization vectors of the incident and scattered light, respectively; **R** – the Raman tensor | |
|---|---|---|---|
| | | Parallel **($\hat{e}_i \parallel \hat{e}_s$)** | Perpendicular **($\hat{e}_i \perp \hat{e}_s$)** |
| $A_g$ | $\begin{pmatrix} a & 0 & 0 \\ 0 & a & 0 \\ 0 & 0 & a \end{pmatrix}$ | $a^2$ | 0 |
| $E_g$ | $\begin{pmatrix} b & 0 & 0 \\ 0 & b & 0 \\ 0 & 0 & -2b \end{pmatrix} \begin{pmatrix} -\sqrt{3}b & 0 & 0 \\ 0 & \sqrt{3}b & 0 \\ 0 & 0 & 0 \end{pmatrix}$ | $b^2(1 + 3\cos^2 2(\theta + \varphi))$ | $b^2 \sin^2 2(\theta + \varphi)$ |
| $T_g^{(1)}$ $T_g^{(3)}$ $T_g^{(2)}$ | $\begin{pmatrix} 0 & 0 & 0 \\ 0 & 0 & d \\ 0 & e & 0 \end{pmatrix} \begin{pmatrix} 0 & 0 & d \\ 0 & 0 & 0 \\ e & 0 & 0 \end{pmatrix} \begin{pmatrix} 0 & d & 0 \\ e & 0 & 0 \\ 0 & 0 & 0 \end{pmatrix}$ | $d^2(\sin^2 2(\theta + \varphi) + C)$ | $d^2 \cos^2 2(\theta + \varphi)$ |

The angular dependence of the Raman scattering intensity for each phonon mode is presented in **Figure 3**. The experimental data were fitted using analytical expressions derived from the Raman tensors for cubic $MnS_2$ (space group $Pa\bar{3}$), [13] assuming all tensor elements to be real, as listed in **Table 1**.

The measured intensity variations are generally well reproduced by the model. In the perpendicular polarization configuration, the intensities of all three $T_g$ modes vanish at the same polarization angles, in agreement with the predictions of the Raman tensors. In contrast, the behavior observed in the parallel polarization configuration differs for the highest-frequency $T_g$ mode, $T_g^{(2)}$. Whereas the intensities of the $T_g^{(1)}$ and $T_g^{(3)}$ modes decrease to nearly zero at specific angles, the intensity of $T_g^{(2)}$ remains finite throughout the entire angular range. To reproduce this behavior, an additive constant was included in the fitting function for this mode. A similar deviation from the idealized tensor model was observed for the $A_g$ mode. Although this mode exhibits its highest intensity in the parallel polarization configuration, a weak but clearly detectable signal is also present in the perpendicular configuration, where the ideal Raman tensor predicts zero intensity.

The residual intensity observed for the $A_g$ and $T_g^{(2)}$ modes suggests that the polarization dependence of the Raman scattering cannot be fully described within the framework of real-valued Raman tensors. One possible explanation is coupling between the $A_g$ and $T_g^{(2)}$ modes. Given their relatively small frequency separation (approximately. 7 $cm^{-1}$) and the fact that both vibrations involve stretching motions of the Te–Te dimers, energy transfer between the two oscillators may occur, preventing the intensity of the $T_g^{(2)}$ mode from reaching zero.

A more plausible explanation involves complex-valued Raman tensors. Such behavior may arise in the presence of resonant electron–phonon coupling and strong optical absorption, which can introduce substantial phase differences between the tensor elements (e.g. [29]). Several features of the experimental polar plots appear to support this scenario. In particular, the lobes of the $E_g$ mode exhibit a noticeably flattened shape compared to the idealized angular dependence predicted for real-valued tensors. Furthermore, the $A_g$ and $T_g^{(2)}$ modes retain finite intensity in the perpendicular and parallel polarization configurations, respectively.

The latter observation is especially remarkable for the $T_g^{(2)}$ mode. Within the Raman-tensor formalism, a non-vanishing intensity in the parallel configuration requires the tensor elements *d* and *e* to differ, as may occur under resonant excitation conditions. A similar possibility was previously discussed by Vogt et al. [13]. If this interpretation is correct, resonance effects could also account for the distinct polarization responses of the three $T_g$ modes allowed by the $Pa\overline{3}$ crystal symmetry. Although mode coupling between the $A_g$ and $T_g^{(2)}$ vibrations cannot be excluded, the excitation-energy dependence of the Raman spectra discussed below lends further support to the interpretation based on resonance-induced complex Raman tensors.

**Excitation-dependent measurements**

The Raman scattering response of $MnTe_2$ was found to depend strongly on the excitation energy. **Figure 4(a)** compares Raman spectra acquired using excitation energies of 1.96 and 2.54 eV. The lower inset presents an enlarged view of the low-frequency spectral region, whereas the upper inset shows a detailed view of the spectrum recorded using the 488 nm laser line. Broad spectral features observed in the 120–180 $cm^{-1}$ range and attributed to crystalline and amorphous tellurium are denoted as c-Te and a-Te, respectively.

The measurements were performed using comparable laser powers of approximately 140 µW. Under these conditions, the overall Raman intensity obtained with 1.96 eV excitation was approximately four times greater than that measured with 2.54 eV excitation. In addition to the increase in total scattering intensity, a pronounced enhancement of the $A_g$ and $E_g$ modes was observed for the lower-energy excitation, enabling the detection of a second-order feature at approximately 360 $cm^{-1}$, which is tentatively assigned to the 2Ag overtone. In contrast, the $E_g$ mode was barely detectable under 2.54 eV excitation, even in the parallel polarization configuration. Similar trends, including changes in the relative phonon intensities and the appearance of the overtone signal, were observed when comparing spectra acquired with excitation energies of 1.58 eV (785 nm) and 2.41 eV (514.5 nm), as shown in **Figure 4(c)**. These observations suggest that excitation in the lower-energy (longer-wavelength) spectral range approaches resonance with an optical transition in $MnTe_2$, leading to a selective enhancement of the Raman scattering cross-sections of the $A_g$ and $E_g$ modes.

Further evidence for the role of resonance effects is provided by the polarization dependence of the $T_g^{(2)}$ mode. For excitation at 2.54 eV, its angular dependence is adequately described by the conventional Raman tensor with real-valued elements and $d = e$. As shown in **Figure 4(b)**, the intensity of the $T_g^{(2)}$ mode decreases to the background level at specific polarization angles, in the same manner as observed for the $T_g^{(1)}$ and $T_g^{(3)}$ modes (not shown). Thus, the anomalous non-vanishing intensity of the $T_g^{(2)}$ mode observed under lower-energy excitation is absent at higher excitation energies. This finding is consistent with the interpretation that resonance-enhanced optical absorption modifies the Raman tensor elements, thereby differentiating the polarization response of the three $T_g$ modes.

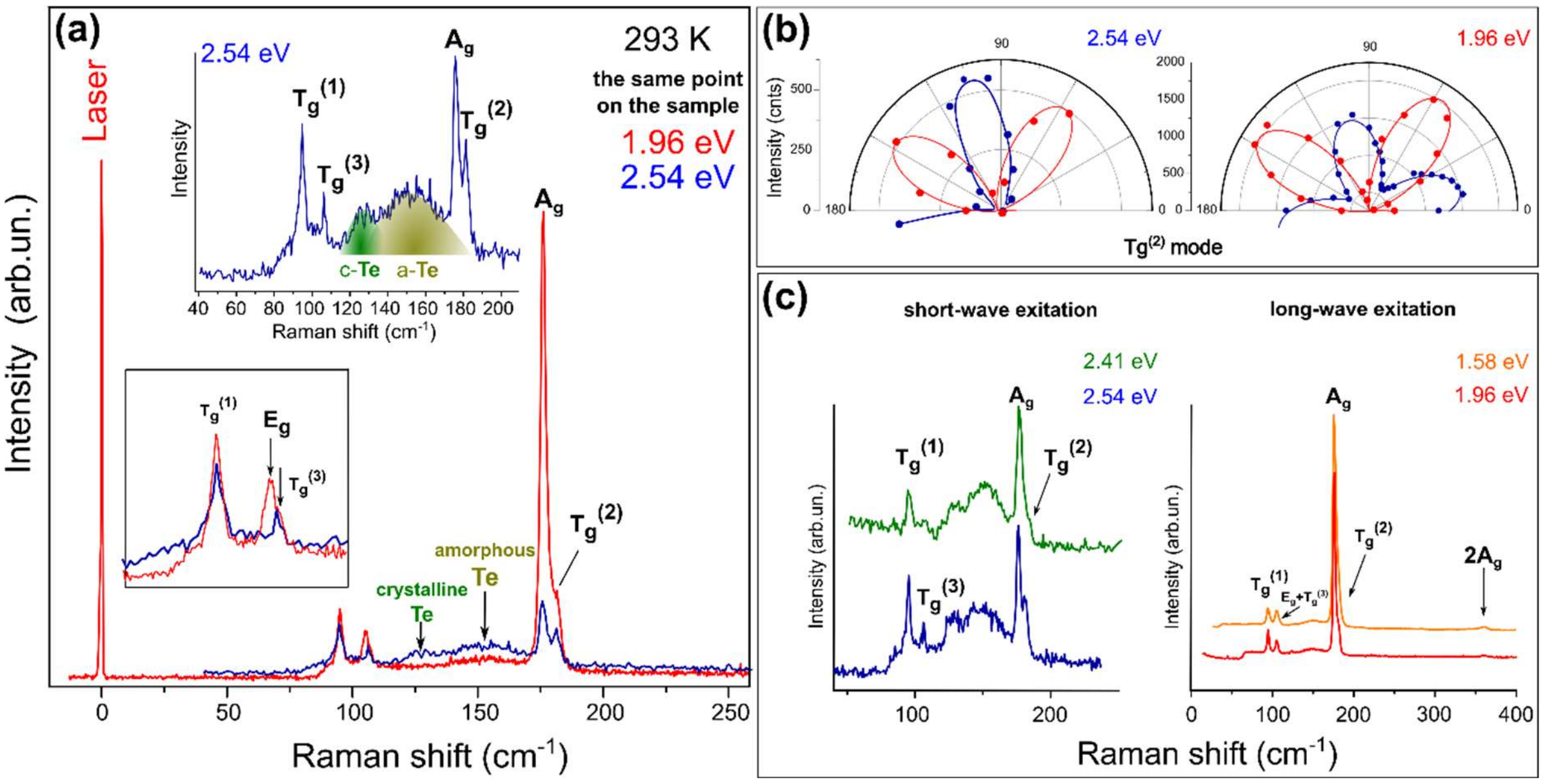

**Figure 4.** Raman scattering in $MnTe_2$ excited by different wavelengths: **(a)** Polarized Raman spectra excited by 488 nm (2.54 eV) and 633 nm (1.96 eV) are presented as blue and red solid lines, respectively. **(b)** Angular dependences of scattering intensity of $T_g^{(2)}$ mode plotted for 1.96 eV and 2.54 eV excitation energy. Data shown in blue and red were collected for parallel and perpendicular measurement configuration, respectively. **(c)** Non-polarized spectra acquired with short- and long-wave excitation are shown for comparison.

As regards tellurium precipitates, Raman scattering is a very effective tool to detect them owing to the high Raman cross-section of c-Te. Using this method such precipitates were reported in numerous tellurides (both in bulk crystals and 2D layers) [30] including MnTe [31]. In our measurements it was observed that tellurium being exposed to laser (regardless of the wavelength) underwent transformation from a-Te into the trigonal c-Te. Furthermore, increasing signal of spectral bands corresponding to crystalline Te was observed with time. Kinetics of the transformation were found to depend strongly on laser power. A considerably higher signal was observed in the case of short-wave (i.e. high energy) excitation (Figure 4 (c)) for which one would expect smaller penetration depths if compared to the long wavelength light. At the moment, the source of tellurium in the evaluated sample is unclear: whether the excess amount of a-Te is deposited during crystal growth process or this is just a result of laser-induced partial decomposition of the sample. One can also suppose that at least some of the numerous reported observations of c-Te precipitates in tellurides by Raman scattering reported in the literature could be due to crystallization of a-Te caused by laser heating.

**Temperature-dependent measurements**

Comparison of Raman spectra taken at room temperature and at 89 K is shown in **Figure 5**. The same set of five signals related to Raman-active phonon modes of $MnTe_2$ are present in both spectra, although signals at 89 K are shifted to higher frequencies and better resolved due to clear narrowing of band widths. In **Figure 6** we plotted experimental data showing the detailed change of phonon frequencies and line widths in the range of temperatures 80 – 343 K. Upon cooling, a gradual and linear increase of vibration frequencies for all five Raman modes was observed in the whole temperature range studied. The observed linear behavior in the positions of Raman peaks with temperature could indicate an anharmonic decay via three-phonon processes well described by the simple Klemens model [32]:

$$\omega(T) = \omega(0) - A\left[1 - \frac{2}{e^{h\omega_0/2k_BT} - 1}\right]$$

where A – anharmonic constant.

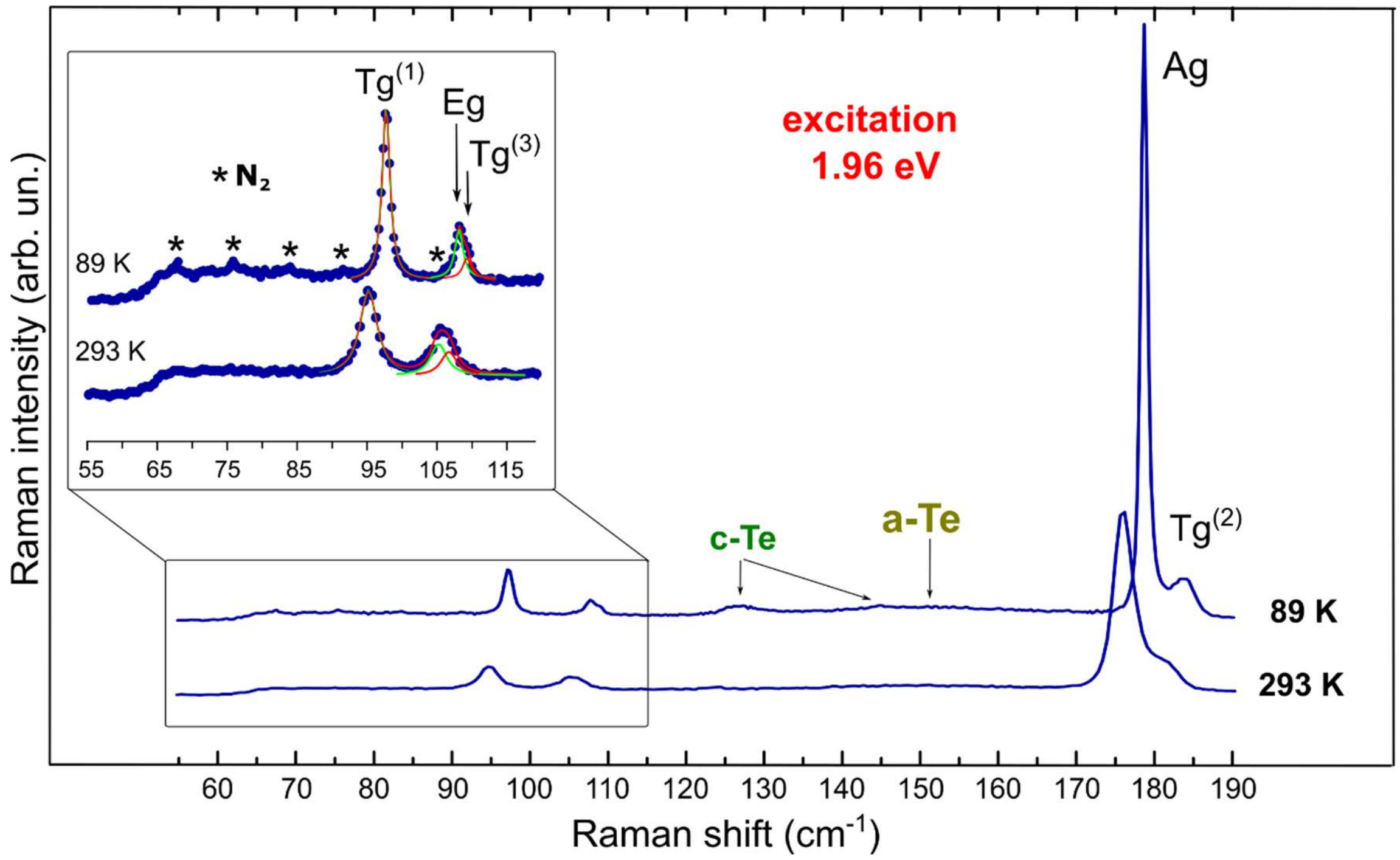

**Figure 5.** Raman spectra of paramagnetic $MnTe_2$ at 293 K and 89 K (blue lines offset vertically for clarity). The low-frequency spectral range is shown in the inset. Lorentzian fits of the bands corresponding to $Tg^{(1)}$, Eg and $Tg^{(3)}$ modes are shown as green and red lines. Asterisks indicate signals from nitrogen used as a cooling agent.

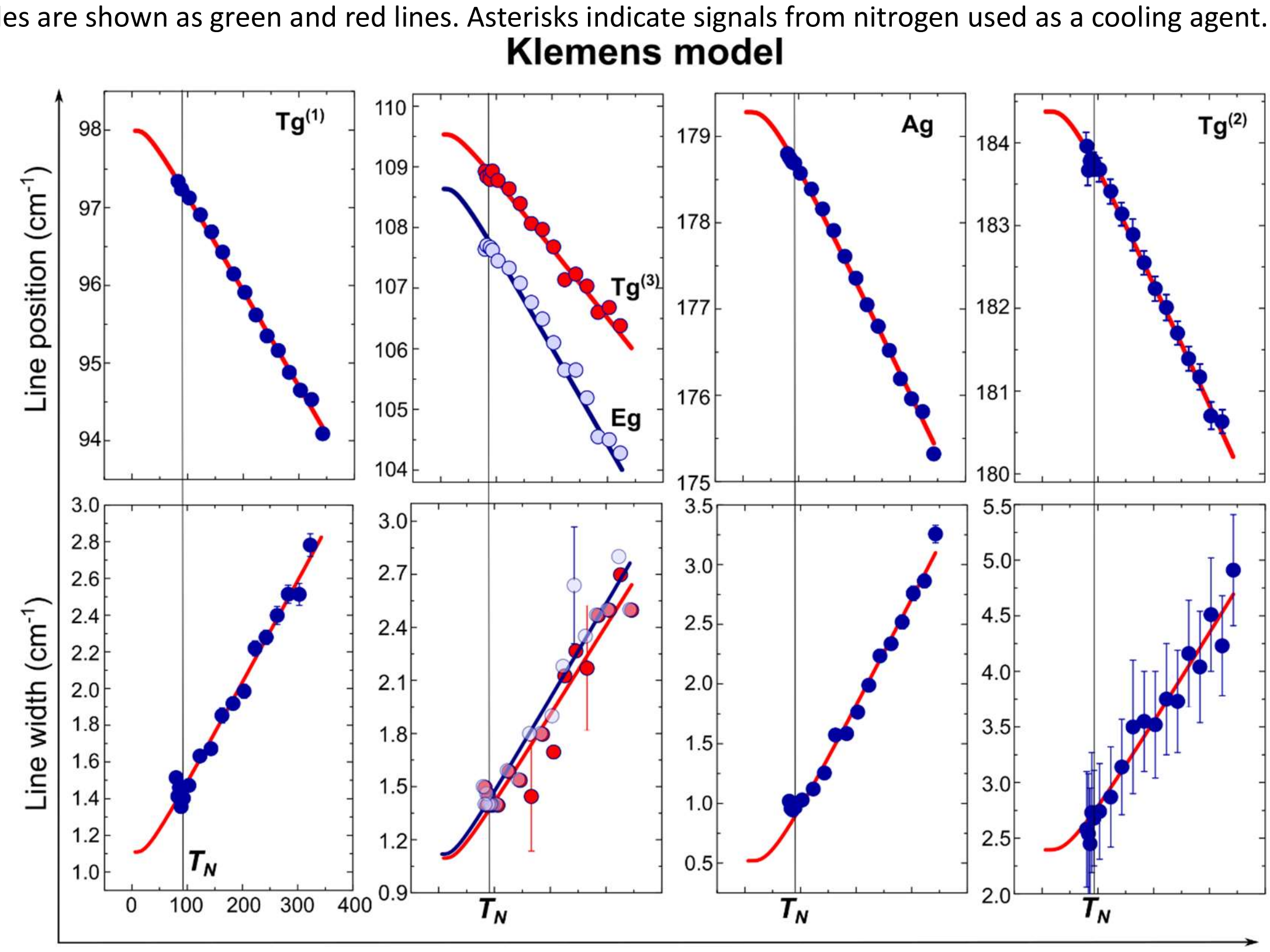

**Figure 6. Upper row:** experimental data of the mode frequencies (solid circles) measured upon lowering temperature displayed along with fits to the Klemens model (solid lines). Error bars are shown for $T_g^{(2)}$ mode whereas for other modes error bars are less than symbol size. "$T_N$" stands for Néel temperature, which equals to 90 K. **Lower row:** changes in measured line widths (FWHM) of corresponding modes of $MnTe_2$ are shown as solid symbols and solid lines for experimental data and fits to Klemens model, respectively.

Line width was fixed while fitting the shape of $E_g$ and $T_g^{(3)}$ bands with Lorentzian profiles, except for spectra collected at 243 K ($E_g$), 163 K and 263 K ($T_g^{(3)}$), for which linewidth of specific mode was left to vary during the fit. Consequently, error bars are shown for the cases of unconstrained fit.

Notably, temperature coefficients of phonon frequencies for $A_g$ and $T_g^{(2)}$ modes showed close values ($7.9 \cdot 10^{-3}$ and $8.9 \cdot 10^{-3}$ cm$^{-1}$/K, respectively) and were considerably higher if compared to values found for $T_g^{(1)}$, $E_g$ and $T_g^{(3)}$ modes, ca. $5.3 \cdot 10^{-3}$ cm$^{-1}$/K. Also, it is worth noting that spectral separation between $E_g$ and $T_g^{(3)}$ is gradually reduced with lowering temperature: from 2 cm$^{-1}$ at 343 K to ca. 1 cm$^{-1}$ at 80 K. At the same time, linewidths experience linear narrowing with temperature (Figure 6 (upper row)) when cooling to ca. 90 K. Further temperature decrease (from 90 K to 80 K) leads to the change of the trend from narrowing to discernible broadening of $T_g^{(1)}$, $E_g$, $T_g^{(3)}$ and $A_g$ lines (Figure 6 (lower row)). This observation might be a tentative indication of the incipient macroscopic magnetic ordering below 90 K, which we observed in the $MnTe_2$ sample from the same batch by the ex-situ measurements in SQUID (see Supplementary Information **Figure S2**).

## III. RESULTS OF CALCULATIONS

Following the identification of all five Raman-active phonon modes in our measurements performed near the Néel temperature, we extended the study by carrying out DFT calculations for the magnetically ordered phase of $MnTe_2$. The results were subsequently compared with the recently reported, but not yet peer-reviewed, data presented in Refs. [23], [24].

Different DFT approaches were benchmarked by comparing experimentally determined structural and magnetic parameters of $MnTe_2$, including the lattice constant, Te–Te and Mn–Te bond lengths, and the magnetic moment, with the corresponding calculated values obtained using the r$^2$SCAN, PBE+U, and PBE$_{sol}$+U functionals. As shown in **Figure S4** in the Supporting Information, the PBE$_{sol}$+U functional with $U_{eff}$ values between 3.6 and 4.2 eV provides the best overall agreement with experiment, consistent with previous studies [9]. Consequently, all theoretical results discussed below are based on PBEsol+U calculations.

According to the neutron diffraction study of Burlet et al. [33], the magnetic ground state of $MnTe_2$ is a non-collinear antiferromagnetic configuration in which the magnetic moments of the four Mn ions within the cubic unit cell point along different body diagonals. Our calculations confirm that this arrangement, denoted S0 in **Figure 7**, corresponds to the lowest-energy magnetic solution. We note that although this state remains as the lowest-energy one at the PBE+U level of theory, the r$^2$SCAN functional incorrectly predicts the S1 arrangement as the ground state.

The S0 ground state is followed by the nearly degenerate S1 and S4 configurations, which lie only 1.6–1.9 meV per $MnTe_2$ formula unit higher in energy, depending on the value of $U_{eff}$ (**Figure 7**). Both configurations may be viewed as ferromagnetic (001) planes stacked antiferromagnetically along the c axis. The distinction between them lies in the orientation of the magnetic moments: in the S4 state they are confined to the (001) plane, whereas in the S1 state they are directed along two body diagonals of the cubic unit cell.

The next-lowest-energy configuration is the S5 state, located 2.2–2.5 meV above the ground state, in which the magnetic moments are aligned along diagonals of the (001) planes. This is followed by the S2 state, which lies 3.3–3.8 meV above S0. Interestingly, the calculated energy differences between the S0 and S2/S5 configurations are of the same order as the zone-center magnon energies reported recently for $MnTe_2$ (2.5 ± 0.2 meV and 4.1 ± 0.5 meV at zero magnetic field) [4]. In contrast, the ferromagnetic configuration (S3) is considerably less stable, lying 54.7–68.3 meV above the ground state.

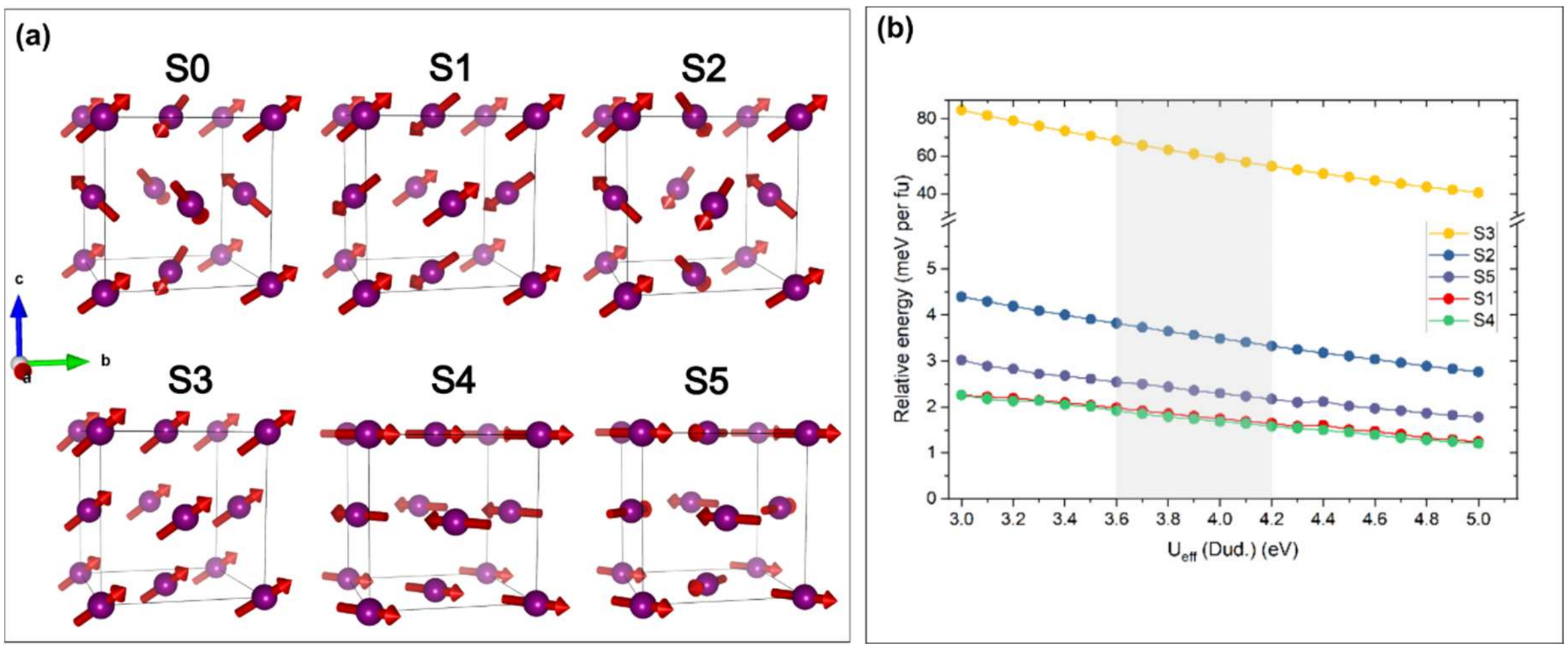

**Figure 7. (a)** Possible spin states of $MnTe_2$, and **(b)** their relative energies (calculated with respect to the S0 state) computed for various $U_{eff}$ values with the $PBE_{sol}$+U method.

The calculated Raman spectrum of the S0 ground state contains five Raman-active phonon modes with the same symmetries as those identified experimentally. **Table 2** compares the phonon frequencies obtained in the present work with the low-temperature experimental values reported in Ref. [23] The $A_g$ and $T_g^{(2)}$ modes are dominated by stretching vibrations of the $Te_2^{2-}$ dimer. Because the $PBE_{sol}$+U functional slightly overestimates the Te–Te bond length, the corresponding phonon frequencies are underestimated by approximately 9% for $U_{eff}$ values between 3.6 and 3.9 eV. In contrast, the $T_g^{(1)}$, $T_g^{(3)}$, and $E_g$ modes primarily involve librational motions of the $Te_2^{2-}$ units and distortions of the $MnTe_6^{4-}$ octahedra (see **Figure S5** in the Supporting Information). Their frequencies are reproduced considerably better, with calculated deviations ranging from −2% to −6%, largely independent of $U_{eff}$.

**Table 2.** Comparison of the experimental frequency (in $cm^{-1}$) of the Raman active modes of $MnTe_2$ compared with DFT calculations

| **Mode** | **Frequency** | |
|---|---|---|
| | **$PBE_{sol}$+U ($U_{eff}$ = 3.8 eV) (This study)** | **EXP (T = 1.6 K) [23]** |
| $T_g^{(2)}$ | 167.4 | 183.1 ± 1.6 |
| $A_g$ | 162.8 | 179.1 ± 0.8 |
| $T_g^{(3)}$ | 105.9 | 107.3 ± 1.6 |
| $E_g$ | 105.4 | |
| $T_g^{(1)}$ | 92.8 | 97.6 ± 1.6 |

In Ref. [23], the Raman band observed at 107.3 $cm^{-1}$ was assigned exclusively to the $E_g$ mode, and its low-temperature splitting was attributed to the loss of time-reversal symmetry accompanying magnetic ordering. Our calculations for magnetically ordered $MnTe_2$ suggest a different interpretation. Specifically, the $E_g$ and $T_g^{(3)}$ modes are predicted to occur at nearly identical frequencies (105.4 and 105.9 $cm^{-1}$, respectively). This is in line with the small separation between these modes, found in our measurements and also reported in Ref. [24]. Consequently, the experimentally observed splitting is more plausibly explained by the progressive narrowing of two closely spaced phonon bands upon cooling, allowing them to be resolved within the experimental spectral resolution.

## IV. CONCLUSIONS

This is the first comprehensive account of the lattice dynamics in the paramagnetic phase of $MnTe_2$ studied by inelastic scattering of coherent visible light through polarization, excitation energy and temperature dependent measurements. Consequently, a full experimental description of all five Raman-active lattice modes has been presented in accord with the group theory analysis for the pyrite-type crystal structure. The key observation in the non-magnetic phase is the nearly identical energy of two lattice modes with different symmetry, $E_g$ and $T_g$, revealed at all temperatures above the magnetic phase transition. Remarkably, a similar narrow splitting is predicted in our 0 K DFT assessment of phonon frequencies in antiferromagnetic phase of $MnTe_2$. Moreover, calculations reveal several competing antiferromagnetic spin configurations within only a few meV of the magnetic ground state.

If degeneration of $E_g$ and $T_g$ modes in magnetic phase is experimentally verified, our finding will indicate that $MnTe_2$ remains strongly correlated system even in its paramagnetic state having crucial consequences for disentangling mechanisms of lattice and spin wave excitations coupling in this material. This combined experimental and theoretical study constitutes solid reference for further exploration of the interplay between crystal and magnetic orders in unconventional magnet $MnTe_2$.

## V. METHODS

**Sample description and details of synthesis.** Single crystals of $MnTe_2$ were grown using the chemical vapor transport (CVT) method. Stoichiometric amounts of high-purity manganese (Mn, 3N8, Koch-Light Laboratories) and tellurium (Te, 5N, Koch-Light Laboratories) were weighed and loaded into a quartz ampoule along with iodine (POCH, 99.8%) as the transport agent, in a concentration of 3 mg/cm³ of the ampoule volume. The ampoule was then evacuated and sealed under high vacuum. The sealed ampoule was placed in a three-zone horizontal furnace to precisely control the temperature gradient and ensure a flat crystallization zone. The growth process was conducted with the source zone maintained at 600 °C and the crystallization zone at 540 °C. This temperature gradient was sustained for 14 days to facilitate the chemical vapor transport and crystal growth. After the growth period, the furnace was switched off and allowed to cool down to room temperature. To reduce residual tellurium from the surface of the as-grown crystals, selected crystals were annealed under vacuum at 300 °C for 3 hours.

**SQUID characterization.** Magnetometry characterization was performed on a single crystalline sample using Quantum Design Magnetic Measurement Property System (MPMS-7T). Experiments consisted of the measurement of the sample's longitudinal magnetic moment vs. temperature and vs. magnetic fields, in the temperature range 2 K $< T <$ 300 K and magnetic fields up to 7T.
Measurements of magnetization vs. temperature were performed utilizing two protocols: zero field cooled (ZFC): the sample is cooled at zero magnetic field; The field is then applied, and the sample is heated during measurement) and field-cooled on cooling (FCC): the sample is cooled in the presence of magnetic fields, while magnetization is recorded). Isothermal magnetization hysteresis loops were carried out at temperatures of 2, 10 , 50, 100 and 250 K.

**Raman spectroscopy study.** Raman scattering measurements were performed on experimental setups using three Raman spectrometers: Raman spectrometers #1 and #2 (LabRAM Aramis (Horiba Jobin Yvon) and Raman spectrometer #3 (Renishaw inVia Qontor). only single crystalline samples of $MnTe_2$ were used for study. If otherwise was clearly stated, measurements were performed on the same sample piece.
**Polarization-dependent measurements.** Raman spectra were measured in backscattering geometry with parallel ($\hat{e}_i \parallel \hat{e}_s$) and crossed ($\hat{e}_i \perp \hat{e}_s$) configurations, where $\hat{e}_i$ and $\hat{e}_s$ are polarization vectors of incident and scattered light, respectively. For the measurements the natural (100) face of the crystal was used. The sample was rotated in order to measure angle dependence of Raman scattering intensity. Sample was excited by 633 nm and 488 nm of He-Ne laser (Melles-Griot) and solid-state laser

Saphire SF (Coherent), respectively. The laser power was attenuated to approx. 140 µW. Incident laser beam was focused on the probed sample using 100x objective (Olympus). Scattered light was collected by the same objective and transferred to Raman spectrometer #1 equipped with diffraction grating 2400 l/cm and charge coupled device detector (Synapse (Horiba Jobin Yvon)) yielding spectral resolution ca. 0.5 $cm^{-1}$. Stokes parts of Raman spectra were measured in the range from 90 to 250 $cm^{-1}$. Acquisition time was 60 s.

**Temperature-dependent measurements.** Nonpolarized Raman scattering spectra were collected in backscattering geometry from the sample inside a thermally stabilized chamber (Linkam DSC600) in the range of temperature from 80 to 343 K. Sample was excited by 633 nm of He-Ne laser (Melles-Griot). The laser power was attenuated to ca. 0.8 mW. Incident laser beam was focused on the probed sample using 50x long working distance objective (Olympus). Scattered light was collected by the same objective and transferred to Raman spectrometer #2 equipped with diffraction grating 2400 l/cm and charge coupled device detector (Synapse (Horiba Jobin Yvon)) yielding spectral resolution ca. 0.5 $cm^{-1}$. Stokes parts of Raman spectra were measured in the range from 65 to 200 $cm^{-1}$. Acquisition time was 100 s.

**Excitation-dependent measurements.** Nonpolarized Raman scattering spectra were collected from the same sample with crystal face (100) orthogonal with respect to the laser beam. Beside 632.8 nm excitation used for angular- and temperature-dependent measurements, the sample was excited by 488, 514.5 and 785 nm laser lines (Saphire SF (Coherent), $Ar^+$ (Modu-laser) and solid-state diode laser (Renishaw), respectively) coupled with Raman spectrometers #1 (the first laser) and #3 (the other two). Power of the probing lasers was kept well below 1 mW in order to avoid a noticeable buildup of the signal related to elementary tellurium in the measured spectra.

**DFT calculations.** Non-collinear DFT calculations were conducted with the PBE [34] and $PBE_{sol}$ [35] GGA functionals with the inclusion of spin-orbit coupling. In both cases the Coulomb repulsion correction (+U method) was applied acting only on to the *4d* electrons of Mn, The +U method followed the simplified formulation of Dudarev et al. [36] in which only $U_{eff} = U - J$ plays a role. We varied $U_{eff}$ between 3 and 5 eV with $J = 0.8$ eV. Additionally, we performed calculations utilizing the $r^2$SCAN meta-GGA functional [37]. Modelling with the hybrid HSE06 functional proved to be computationally too demanding. Geometry optimization was performed for the lowest-energy spin state (S0 in **Figure 7**). Calculations were performed in VASP 6.4.3 with a cutoff energy of the plane waves was equal to 800 eV, a self-consistent field convergence criterion of $10^{-7}$ eV, and a Monkhorst–Pack 9 × 9 × 9 k-point mesh. Structures were optimized using a conjugate gradient algorithm until forces acting on the atoms were less than 5 meV/Å.

**Phonon calculations** were conducted on the conventional unit cell of $MnTe_2$ with the finite-displacement method as implemented in Phonopy [38] and the frequencies of the Raman-active Γ-point modes were compared with experimental data from Ref. [23]. Phonon dispersions and density of states along with visualizations of atom displacements for particular lattice vibrations are shown in Supplementary Information **Figure S5.**

**Acknowledgements**

D.K., M.H.D. gratefully acknowledge Polish high-performance computing infrastructure PLGrid (HPC Center: ACK Cyfronet AGH) for providing computer facilities and support within computational grant no. PLG/2025/018490. Synthesis of samples for the study was supported by the “MagTop” project (FENG.02.01-IP.05-0028/23) carried out within the “International Research Agendas” programme of the Foundation for Polish Science, co-financed by the European Union under the European Funds for Smart Economy 2021-2027 (FENG). A.S.W. acknowledges the support of the National Science Centre, Poland (NCN), through the MINIATURA 9 with project No. 2025/09/X/ST3/00809.

**Author Contributions:**
**Taras Palasyuk:** conceptualization, methodology, investigation, formal analysis, data curation, project administration, resources, supervision, validation, visualization, writing – original draft, writing – review & editing
**Dominik Kurzydłowski:** investigation, formal analysis, data curation, writing – review & editing
**Ashutosh S. Wadge:** resources, writing – review & editing
**Rajmund Bacewicz:** formal analysis, writing – review & editing
**Madhavi H. Dalsaniya:** investigation
**Michał Wierzbicki:** formal analysis
**Daniel Jastrzebski:** resources, writing – review & editing
**Andrzej Wiśniewski:** conceptualization, resources, writing – review & editing
**Wojciech Szuszkiewicz:** writing – review & editing
**Igor Antoniazzi:** investigation, formal analysis, resources, writing – review & editing
**Bruno Camargo:** investigation, formal analysis, resources, writing – review & editing
**Maciej Molas:** resources, writing – review & editing
**Adam Babiński:** resources, writing – review & editing
**Krzysztof Zberecki:** writing – review & editing
**Cezariusz Jastrzebski:** conceptualization, methodology, resources, supervision, investigation, formal analysis, data curation, writing – review & editing